\documentclass{article}

\usepackage{microtype}
\usepackage{graphicx}
\usepackage{subfigure}
\usepackage{booktabs} 

\usepackage{hyperref}
\usepackage{cleveref}
\usepackage[frozencache=true,cachedir=_minted-output]{minted}
\usepackage{url} 
\usepackage{breakurl}

\usepackage[accepted]{icml2020}

\icmltitlerunning{Serverless inferencing on Kubernetes}

\begin{document}

\twocolumn[
\icmltitle{Serverless inferencing on Kubernetes}




\begin{icmlauthorlist}
\icmlauthor{Clive Cox}{seldon}
\icmlauthor{Dan Sun}{bloomberg}
\icmlauthor{Ellis Tarn}{}
\icmlauthor{Animesh Singh}{ibm}
\icmlauthor{Rakesh Kelkar}{microsoft}
\icmlauthor{David Goodwin}{nvidia}
\end{icmlauthorlist}

\icmlaffiliation{seldon}{Seldon Technologies}
\icmlaffiliation{ibm}{IBM}
\icmlaffiliation{nvidia}{NVIDIA}
\icmlaffiliation{bloomberg}{Bloomberg, L.P.}
\icmlaffiliation{microsoft}{Microsoft}

\icmlcorrespondingauthor{Clive Cox}{cc@seldon.io}
\icmlcorrespondingauthor{Dan Sun}{dsun20@bloomberg.net}
\icmlcorrespondingauthor{Ellis Tarn}{ellistarn@gmail.com}

\icmlkeywords{Machine Learning, inference, kubernetes, deployment, ICML}

\vskip 0.3in
]



\printAffiliationsAndNotice{}  

\begin{abstract}
Organisations are increasingly putting machine learning models into production at scale. The increasing popularity of serverless scale-to-zero paradigms presents an opportunity for deploying machine learning models to help mitigate infrastructure costs when many models may not be in continuous use. We will discuss the KFServing project which builds on the KNative serverless paradigm to provide a serverless machine learning inference solution that allows a consistent and simple interface for data scientists to deploy their models. We will show how it solves the challenges of autoscaling GPU based inference and discuss some of the lessons learnt from using it in production.
\end{abstract}

\section{Introduction}\label{sec:intro}
The use of machine learning models in industry which was previously the domain of a few high tech
companies is now being democratised with a range of open frameworks such as MLFlow \cite{mlflow} and Kubeflow \cite{kubeflow} being offered alongside the managed cloud offerings from Google, Amazon and Microsoft amongst others. With more models being put into production the challenges of the deployment and inference stage in the machine learning life-cycle have become more prominent. Machine learning inference at scale can produce a large cost burden with the need for high numbers of CPU and GPU servers to cover the real time inference demands. However, many models have cyclical or sporadic usage patterns so autoscaling as well as scaling to zero becomes an important criteria for any organisation hoping to provide a consistent service at reduced cost. In recent years the emergence of serverless as a paradigm to allow for transient functions to be deployed on demand has become popular \cite{passwater}. In this paper we will discuss the solution provided by the KFServing \cite{kfserving} project at realising the serverless advantages for machine learning and the realities of putting these techniques into production while discussing the open problems that remain.

\section{Challenges of Deployment and Inference}\label{sec:deploy}
When organisations come to deploy their models into production they are presented with a range of challenges. We focus on four core challenges:

\begin{enumerate}
    \item Handling multiple machine learning frameworks in a consistent manner.    
    \item Updating running models with new versions.
    \item Scaling models appropriately with constraints.
    \item Monitoring models.
\end{enumerate}

First, different problems require the use of particular machine learning frameworks with a wide variety in popular use including Tensorflow \cite{tensorflow}, PyTorch \cite{pytorch}, XGBoost \cite{xgboost},and  SKlearn \cite{scikit-learn} as well as specialized optimization solutions including RAPIDS \cite{rapids} and Intel\textregistered OpenVino\textsuperscript{TM} \cite{openvino}. Any deployment methodology needs to provide a consistent abstraction across these frameworks to allow solutions to be provided using the most applicable technology but configured in a consistent manner. Even though standardisation projects such as ONNX \cite{onnx} have gained ground there is still a strong desire for native framework serving solutions to sit alongside ONNX servers especially as ONNX still only covers a subset of describable models.

Secondly, once in production the models being served will most likely need to be updated with improved versions. New models can be evaluated via a range of deployment techniques including canary and shadow deployments in combination with rolling updates and red green strategies. Canaries allow users to split a small percentage of traffic to their new model while shadows allow full duplication of production traffic to the live server alongside a ``shadow" model with only the live server's response being returned to the client. Both techniques allow the stability and correctness of the new model to be ascertained in production  before the decision to do a final deployment via a rolling update gradually replacing old models with new ones or red-green strategies allowing the full new set of deployments to be started and a switch to the new when ready. 

Thirdly, once a model is in production it needs to be scaled to the appropriate level for the concurrency, latency and throughput requirements. It also needs to be reactive to scale up and scale down as the demand placed on it changes over time. These latency and throughput demands need to be traded off against the cost of running the model at scale.

Fourthly, running models need to be monitored and inspected to ensure they are behaving as expected. There are a variety of concerns here. Core metrics such as latency, throughput and errors need to be handled as well as data science specific metrics such as accuracy and input and output distribution analysis. Running models may be prone to drift if the input distribution changes which may adversely affect performance. Individual outliers may need to be detected as well as targeted adversarial attacks. In summary, given models run in uncertain and changing environments and represent powerful functions whose behaviour is rarely fully determined they need to be constantly monitored.

In the next section we will show how current trends in infrastructure provisioning have provided a solution for these challenges.

\section{Machine Learning on Kubernetes}\label{sec:k8s}

Containerized infrastructure has grown in popularity since the emergence of Docker \cite{docker}. Docker allowed one to package a single application and all its dependencies so that the resulting image could be run anywhere. However, the challenge remained of how to build and orchestrate applications made up of many such images and how to run those at scale on many servers sharing resources with a range of applications. The Kubernetes project \cite{kubernetes, kubernetes-overview} was created to solve this requirement and has become the de facto leader in the field allowing a true cluster compute solution for organisations to run their applications on any cloud or on premise as needed. With the Kubernetes base a multitude of focused projects have built on it to provide core services such as logging, metrics, network management, and security and provide a burgeoning Cloud Native \cite{cncf} ecosystem. In the last few years machine learning projects have begun to provide focused solutions for data scientists building on this maturing Kubernetes stack. In the next section we discuss how the KFServing project was created to solve the core challenges outlined earlier building upon a set of Kubernetes projects for a focused machine learning inference solution.

\section{KFServing}\label{sec:kfserving}

\begin{figure}[htbp]
    \centering
    \includegraphics[width=\columnwidth]{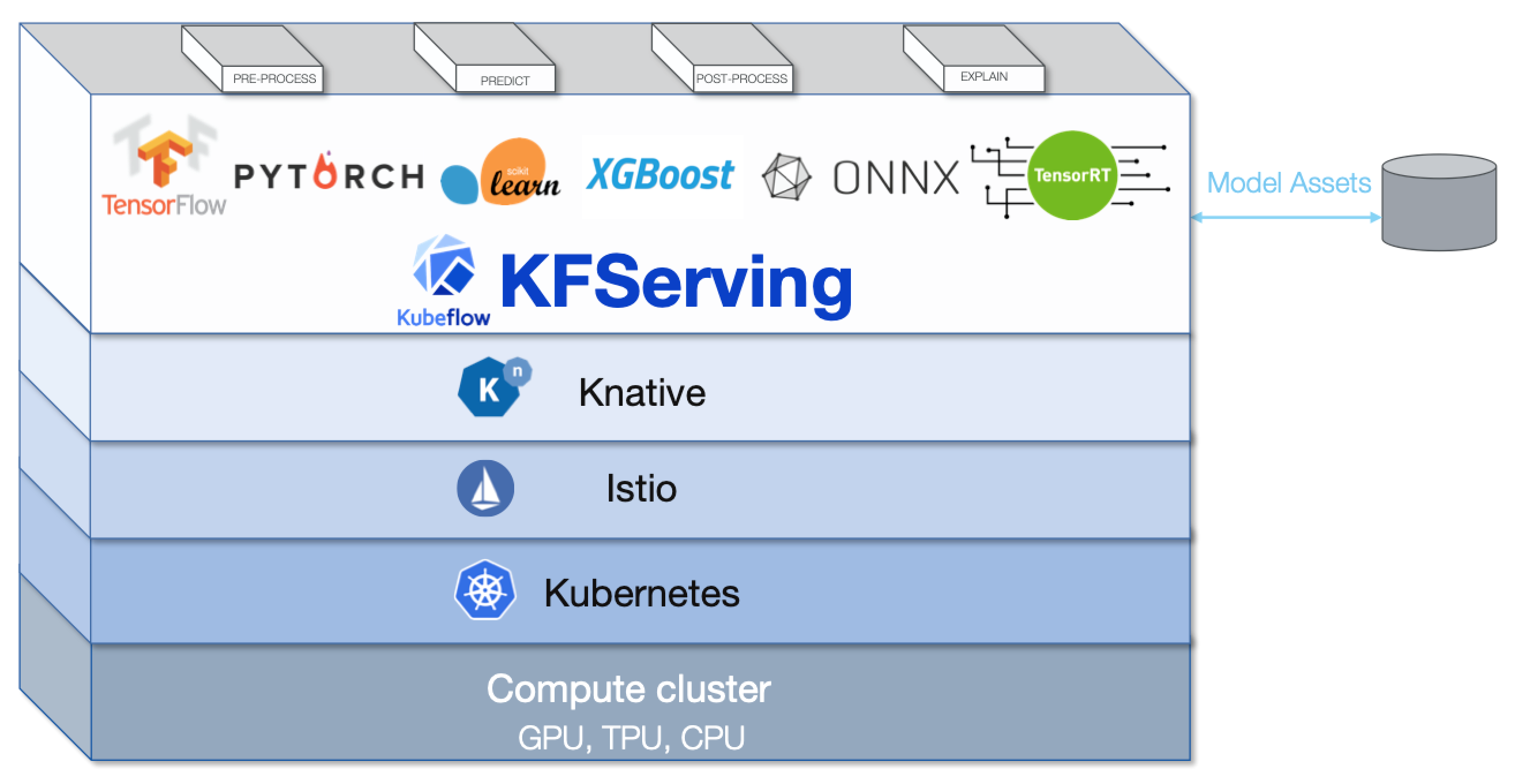}
    \caption{KFServing cloud native technology stack}
    \label{fig:kfserving_stack}
\end{figure}

KFServing is a project that was created within the Kubeflow \cite{kubeflow} project ecosystem. Kubeflow aims to bring a suite of compatible projects for end to end machine learning to Kubernetes. KFServing is focused on building on the serverless KNative \cite{knative} Kubernetes project to extend Kubernetes to allow machine learning inference. KNative itself is built upon service mesh technologies including Istio \cite{istio} which provides routing, security and traffic management. \Cref{fig:kfserving_stack} shows the KFServing technology stack allowing diverse machine learning server technologies to be easily utilized on a Kubernetes cluster containing CPU, GPU and TPU capabilities.

One of the challenges in utilizing Kubernetes is there is a wide range of API and resource types that need to be correctly configured to build a fully functioning application. KFServing intends to provide a simple Custom Resource Definition (CRD) \emph{InferenceService} in which a data scientist can connect their saved model with appropriate deployment technology which is managed for them. An example InferenceService is shown below for a Tensorflow model. 

\begin{minted}
[
breaklines,
fontsize=\footnotesize,
]
{yaml}
apiVersion: serving.kubeflow.org/v1alpha2
kind: InferenceService
metadata:
  name: flowers-sample
spec:
  default:
    predictor:
      tensorflow:
        storageUri: gs://kfserving-samples/models/ tensorflow/flowers
\end{minted}

By simply specifying the type of server and location of the model artifacts KFServing will:

\begin{itemize}
    \item Create a serverless KNative service with the appropriate server image.
    \item Create a storage initializer to download the artifacts from any popular storage (Google Storage, Amazon S3, Azure, local disk) and load onto the server.
    \item Wire up networking so an endpoint is made available for inference requests.
\end{itemize}

Users can further customize their InferenceService to add resource requests for CPU, GPU, TPU and memory requests and limits. Correct scheduling will then take place to locate the model server onto available Kubernetes nodes with the requested resources and ensure limits are not breached.

Model updates can easily be provided by updating the InferenceService with a \emph{canary} section as illustrated below where 10\% of traffic will be sent to the canary model in this example.

\begin{minted}
[
breaklines,
fontsize=\footnotesize,
]
{yaml}
apiVersion: serving.kubeflow.org/v1alpha2
kind: InferenceService
metadata:
  name: flowers-sample
spec:
  default:
    predictor:
      tensorflow:
        storageUri: gs://kfserving-samples/models/ tensorflow/flowers
  canaryTrafficPercent: 10
  canary:
    predictor:
      tensorflow:
        storageUri: gs://kfserving-samples/models/ tensorflow/flowers-2
\end{minted}

The design pattern of Kubernetes is that infrastructure definitions are declarative and new versions of a resource definition force a reconciliation process to change the infrastructure running on the cluster to eventually reflect the current definition. The process allows for the beneficial ``GitOps" \cite{gitops}  pattern to be followed where every version of a resource is committed to source control (e.g., Github) allowing clear audit histories and easy rollback to previous versions.

By utilizing KNative as a backbone, entire model servers can be scaled to zero if there are no incoming requests allowing the cluster to schedule other model services that currently have live requests. 

KFServing further allows a data scientist to add \emph{transformers} and \emph{explainers} to the core model server. Transformers allow focused data transformations of the request and response from the model. For example, a text model may need input words transformed into feature embedding vectors which are the raw input to the model. Explainers allow model explanation methods to be attached to the service so an individual request/response from the model can be sent for providing human understandable explanations. This allows users and auditors to better understand why a model is providing the predictions for certain inputs.

Finally, KFServing allows payload logging to be switched on to send request and response payloads from the model to be asynchronously processed for monitoring and analysis needs.

\subsection{GPU Autoscaling}
Models requiring GPUs at inference time are becoming more common. However, correctly autoscaling models which utilize GPUs is a challenge. GPU duty cycle metrics are only becoming easily available on some Kubernetes Cloud platforms and configuring Horizontal Pod Autoscalers (HPAs) for them is not easy. Furthermore, you need to combine the CPU usage of the running server alongside its GPU usage which is not an easy task to reconcile for creating an autoscaling decision. Other alternatives such as latency based autoscaling can allow scaling up decisions but are harder to implement for scaling down decisions \cite{kaiser_2020}. As KFServing utilizes KNative it can take advantage of the KNative Pod Autoscaler (KPA) which implements request based autoscaling to provide a general solution to autoscaling.

Request based autoscaling looks at how many requests are in-flight (e.g., being served but not yet responded to) and sees what the current available concurrency is (assumed 1 per server but can be configured). By this simple metric a decision can be made to scale up resources or scale down resources as needed without having to delve into complex GPU and CPU metrics. Request based autoscaling therefore provides a generic solution to autoscaling which is intuitively understandable and doesn't require extensive optimization to provide acceptable performance.

\section{Production Experience}\label{sec:production}

KFServing has been running in production at multiple organisations as of early 2020 \cite{gtc, gojek, coreweave}. These deployments have included running challenging GPT-2 \cite{gpt2} models at scale \cite{aidungeon}. From these case studies several learnings have arisen which are discussed below.

The KPA is integral to providing scale up/down capabilities and is partially implemented by including a sidecar container (queue-proxy) running next to your desired service which grabs metrics about the in-flight requests. Quotas are set on how much CPU this sidecar can utilize. However, linux kernel CFS (Completely Fair Scheduler) scheduling bugs \cite{tail-latency} mean that even a well behaved container with mostly IO tasks can be CPU throttled and in the case of KNative this can adversely affect the tail latency for your model server unnecessarily and therefore careful monitoring of the CPU throttling metrics is required.

In some scenarios where latency is at a high premium the time taken to autoscale in serverless scenarios for the initial request (from zero) or for autoscaling up can be prohibitive especially when model artifacts can be large. For these scenarios the benefits of serverless are less apparent when consistent low latency is required for all requests and must be traded off against the infrastructure cost gains.

Batching individual model inference requests is important to unlock the high throughput
when running inference workloads on GPU and most of ML/DL frameworks are optimized for batch requests. However, the user often needs to run enough tests to find the optimal batch size depending on the traffic pattern or the number of models loaded on the GPU. When services are receiving transaction rates per second less than the batch size it will lead to "batch delay" and cause response latency per request to spike. Careful or dynamic tuning is therefore required based on the load pattern.

Last but not least, managing the Istio/Knative stack at scale with hundreds or thousands of inference services can be a challenging task. Every Istio Virtual Service increases the memory footprint on the  Ingress gateways, If you are running a service mesh then every sidecar adds more load to the Istio's control panel, hence regular adjustment of the number of replicas or CPU/memory limits for gateway and control panel deployments are required. Fortunately, Istio version 1.5 has consolidated all the control plane components into a single one called istiod which helps simplify administration overhead for cluster managers.

\section{Open Problems}\label{sec:open}

KFServing provides many advantages as discussed to facilitate data scientists to get their models into production. However, several challenges still exist and need further research.

Many of today's models can be very large, e.g. 5-30G, and can consume 100\% of the provided CPU to serve a single request. Scaling these in a manner which is still not prohibitive to infrastructure costs is an open problem. Downloading a 5-30G model takes a non-trivial time which means autoscaling latency is adversely affected. Some form of caching and artifact sharing is required to scale large models in an effective manner.

Another observed trend in some scenarios is to create many 100s--1000s of small models trained on different subsets of data. To create individual model servers for these models is unviable. Techniques are required to allow model servers to easily share multiple models in a fashion which is transparent to the end user. Models would be scheduled and autoscaled to available underlying servers and transparently sharded as the traffic and load pattern in the cluster changes.

Effectively monitoring running models is an open research area that covers many topics from real time analysis of the input and output data distributions to installing detectors for outliers, drift and adversarial attacks. Deploying these components in an efficient, automated and accurate manner for each model deployed is a challenging task. KFServing has implemented initial solutions for these that allow created detectors and analysis modules to run asynchronously to the main model serving requests. However, more work needs to be done to ensure all monitoring components are managed, versioned and updated for each new release of a model.

\section{Summary}\label{sec:summary}

In this paper we have discussed the challenges faced putting machine learning models into production and the solutions provided by taking a container based serverless route to solving them as expressed in the KFServing project for machine learning inference on Kubernetes. The varied infrastructure and data science challenges for inference mean many open problems still exist. KFServing is an open source project and is working on addressing these challenges. We welcome collaboration from interested parties to help data science move from research into production in the most seamless manner.


\bibliography{references}
\bibliographystyle{icml2020}

\end{document}